\documentclass[12pt,preprint]{aastex}
\pdfoutput=1

\newcommand{\htwo}{H$_{2}$}

\newcommand{\kms}{km~s{$^{-1}$}}

\newcommand{\per}{$^{\rm{-1}}$}

\newcommand{\msol}{M{$_{\odot}$}}
\newcommand{\lsol}{L{$_{\odot}$}}

\newcommand{\oiii}{[O~III]}

\newcommand{\nii}{[N~II]}

\newcommand{\vt}{V$_{T}$}
\newcommand{\vel}{V$_{R}$}

\newcommand{\ngc}{the Ring Nebula}

\newcommand{\mut}{$\mu$}
\newcommand{\masyr}{mas yr$^{-1}$}
\newcommand{\Escale}{mas yr$^{-1}$/\arcsec}

\slugcomment{To appear in the Astronomical Journal}

\shorttitle{Tangential Motions in NGC 6720}
\shortauthors{O'Dell}

\begin{document}

\title{Studies of NGC 6720 with Calibrated HST WFC3 Emission-Line Filter Images--III:Tangential Motions using AstroDrizzle Images\
\footnote{
Based on observations with the NASA/ESA Hubble Space Telescope,
obtained at the Space Telescope Science Institute, which is operated by
the Association of Universities for Research in Astronomy, Inc., under
NASA Contract No. NAS 5-26555.}}

\author{C. R. O'Dell}
\affil{Department of Physics and Astronomy, Vanderbilt University, Box 1807-B, Nashville, TN 37235}

\author{G. J. Ferland}
\affil{Department of Physics and Astronomy, University of Kentucky, Lexington, KY 40506}

\author{W. J. Henney}
\affil{Centro de Radioastronom\'{\i}a y Astrof\'{\i}sica, Universidad Nacional Aut\'onoma de M\'exico, Apartado Postal 3-72,
58090 Morelia, Michaoac\'an, M\'exico}

\and

\author{M. Peimbert}
\affil{Instituto de Astronomia, Universidad Nacional Aut\'onoma de M\'exico, Apdo, Postal 70-264, 04510 M\'exico D. F., M\'exico}

\email{cr.odell@vanderbilt.edu}

\begin{abstract}

We have been able to compare with astrometric precision AstroDrizzle processed images of NGC~6720 (\ngc) made using two cameras on the Hubble Space Telescope. The time difference of the observations was 12.925 yrs. This large time-base allowed determination of tangential velocities of features within this classic planetary nebula. Individual features were measured in \nii\ images as were the dark knots seen in silhouette against background nebular \oiii\ emission. An image magnification and matching technique was also used to test the accuracy of the usual assumption of homologous expansion. We found that homologous expansion does apply, but the rate of expansion is greater along the major axis of the nebula, which is intrinsically larger than the minor axis. We find that the dark knots expand more slowly that the nebular gas, that the distance to the nebula is 720 pc $\pm$30\%, and the dynamic age of \ngc\ is about 4000 yrs. The dynamic age is in agreement with the position of the central star on theoretical curves for stars collapsing from the peak of the Asymptotic Giant Branch to being white dwarfs. 

\end{abstract}

\section{Background and Introduction}
\label{sec:intro}

Using the spatial motion of the material in planetary nebulae (PNe) to understand their 3-D structure and evolutionary processes has been a goal of many studies of these ubiquitous objects. Olin C. Wilson (1950) first established the utility of radial velocity maps of a PN for determining its basic structure, showing that most objects could be explained as expanding ellipsoidal shells. 
Numerous studies since then have exploited this approach as high-velocity resolution spectrographs became more common and as high quantum efficiency CCD detectors replaced photograph emulsions. We now understand how the apparent form of many planetary nebulae (PNe) is the result interacting stellar winds. Slowly moving gas, expelled  as the central star reaches the top of the Asymptotic Giant Branch, is overtaken by a fast wind of gas, expelled as the star begins to heat  up. This is when the central star begins its collapse towards becoming a white dwarf. In the earliest phase the ejected material is mostly neutral and has a small ionized core. The shell may become fully ionized as the central star increases in temperature and remains at high bolometric luminosity. As the luminosity of the central star in ionizing photons drops, the outer parts of the material may again become neutral. From that point on the balance of ionized and neutral gas becomes a subtle balance. The decreasing gas density that results from the nebula's expansion causes the ionized fraction to increase. However, the decreasing luminosity of the star causes the ionized fraction to decrease.These factors mean that the dynamic structure that we see today is the result of multiple processes, rather than the simple ejection of a shell from a collapsing star.  
Nevertheless, it is worthwhile studying the dynamics of the PNe because it allows us to infer what must have been their earlier conditions and to predict their futures. 

\subsection{3-D Structures from Radial Velocities}
\label{sec:RadVel}

Constraining determination of the 3-D structure and dynamics only to the use of line-of-sight (radial) velocities is an important limitation. 
Interpretation of the observed radial velocities usually demands a simplifying assumption about the general nature of the motion. Most often the assumption is of homologous expansion; that is, that the spatial velocity away from the central star scales with the spatial distance. Homologous expansion is like explosive expulsion of particles of varying velocity. Knowing now that the expanding material is shaped and accelerated by multiple processes makes it surprising that PNe resemble homologous expansion, a result going back to Wilson's (1950) early study. Wilson showed that emission originating farther from the central star shows a higher velocity of expansion. 
Mapping the radial velocities of PNe has become straightforward and common, but these data only map one component of the spatial motion (\vel). However, one cannot underestimate the importance of Wilson's (1950) study since quantitative efforts to determine the 3-D structure of PNe originate with this paper.

\subsection{Determining Tangential Velocities in Planetary Nebulae}
\label{sec:intromut}

Potentially one can create better models for the PNe if we also have measurements of their tangential motions. However, for a fixed value of the spatial motion perpendicular to the line-of-sight (\vt) the apparent motion in the plane of the sky (\mut) decreases with increasing distance (D) to the target, and many fewer nearby candidate objects exist.  When expressed in the common units of \vt~ in kilometers/sec, \mut~ in \arcsec /yr, and D in parsecs, the relation is \mut ~= 0.211$\times$\vt / D. Inserting representative values of \vt~= 20 \kms\ and D = 500 pc gives an expected \mut\ of 0.0089 \arcsec / yr , 8.9 milliarcsec (mas) per year.  Determination of \mut\ was first attempted by Liller et al. (1966) using ground-based images of planetary nebulae obtained with several large telescopes over a period of up to 62 years. No believable values were produced because of the blurring by astronomical seeing (characteristically about 2\arcsec\ in the older images employed) and the lack of a clear filtering of single emission-lines. The latter limitation is because different emission-lines can arise from different positions in the nebula, thus blurring the image.  

As the Very Large Array (VLA) radio interferometer began to produce stable images of 1\arcsec\ resolution and better, the possibility was recognized that this telescope could provide usable \mut\ values by special processing of the data \citep{mas86,gomez93,hajian93,hajian95}. A summary of results was given by Terzian (1997). Limited to the study of high surface-brightness PNe and unable to give unambiguous results for different portions of the PNe, the method is no longer in wide use.

The VLA method of obtaining \mut\ has essentially been succeeded by the use of two-epoch images using the Hubble Space Telescope (HST).
With its stable optical-wavelength images of slightly better than 0.1\arcsec\ resolution and its ability to clearly isolate regions emitting specific emission-lines, this becomes the more promising approach. Reed et~al. (1999) employed pairs of images of NGC~6543 using the high resolution CCD camera  (45.5 mas pixels) of the Wide Field Planetary Camera 2 (WFPC2) with a time separation of 2.92 yrs. They identified three approaches. The first approach was to measure radial profiles of samples on the two sets of images and then look for displacements using a gradient recognition algorithm. The second approach also used radial profiles, but fit gaussian profiles to features and looked for the changes in their centers. The final technique was to magnify the first epoch images by a succession of scale factors, align their centers on the central star, and recognize at what scale factor the magnified first epoch and the second epoch images matched. This last approach utilizes all of the nebula's image, but its accuracy is difficult to quantify and works only in the case of near-homologous expansion. Motions of up to 6.5 \masyr\ were found.  In a subsequent study \citep{pal02} of four PNe with time differences of 3.9 yrs to 4.2 yrs, the first and second approach were used to successfully determine \mut\ in three objects. However, they compared broad bandpass images (the WFPC2 F555W) with monochromatic images (the WFPC2 F502N), arguing that the signal of both filters was dominated by the [O~III] emission line at 500.7 nm.  Values up to 2 \masyr\ were found.

\subsection{Tangential Velocities in the Ring Nebula}
\label{sec:6720history}

As the archetypical elliptical-form  PN, NGC~6720 (henceforth \ngc ) has attracted considerable attention in observational programs. However, it is significantly lower in surface-brightness than most of the PNe that have been the subject of radial velocity and tangential motion studies, and high resolution spectroscopy has been a relatively recent subject. O'Dell et al. (2002) used WFPC2  \nii\ emission-line images with a 1.944 yr time difference and the magnification technique of Reed et al. (1999) to determine a magnification factor (M) of M=1.00067$\pm$0.0001 yr\per. This study employed the coarser (100 mas) pixels of the wider field of view CCD's of the WFPC2.  
 With an average diameter of its Main Ring of 36\arcsec, this corresponds to an expansion of about 24 \masyr. 
 
In a discussion of possibly homologous expansion it is more convenient to express the results in terms of what we call here the expansion scale factor E, where E = \mut /$\phi$, where $\phi$ is the angular distance from a point in the nebula to the central star. If homologous expansion applies, then E will be constant. It is analogous to the Hubble constant that used to describe to the first order of accuracy the expansion of the universe. The expansion scale factor is most conveniently expressed with units of \Escale. For the magnification technique E=(M-1)$\times$10$^{3}$.   The O'Dell et al. (2002) results can then be summarized as E=0.67$\pm$0.1 \Escale.

However, a study of radial velocities in the Ring Nebula \citep{OSH} questioned the accuracy of this result because the second epoch images had the nebular centered on the central star, which meant that the image of the Main Ring fell across all four CCDs in the WFPC2, whereas the first epoch images fell onto only a single CCD.  Since the relative positions of the CCDs are more difficult to determine than the astrometric corrections within a single CCD, there could be a systematic error. This limitation was circumvented in a later study \citep{OHS}, where single CCD images with a time difference of 9.557 yr were employed. All of the techniques introduced by Reed et al. (1999) were used with the result that
E=0.23$\pm$0.1 \Escale\ for \nii\ emission and four dark features seen in F502N gave E=0.30$\pm$0.17 \Escale. These results indicate that the O'Dell et al.\@ (2002) numbers were inaccurate, probably because of using multiple CCDs in the second epoch images. The O'Dell et al. (2009) result also argues for the first time that the dark knots that mark \ngc\ and are ubiquitous in nearby PNe \citep{ode02} expand at approximately the same rate as the nebular material. The big uncertainty in the values of E means that a conclusion about commonality of motion is quite uncertain.

\subsection{3-D Modeling  and the Value of Tangential Velocities in the Ring Nebula}
\label{sec:3Dhistory}

Attempts to model \ngc\ in 3-D have a long history. These are summarized in a recent publication \citep{ode13a} that draws on the extensive imaging and radial velocity mapping that  has been done. In that study, it was found that the Main Ring of the nebula, where most of its optical light originates, is a non-circular disk of ionization bounded gas. Near the central axis of the disk material extends  towards the observer in the form of lobes, seen only in projection onto the plane of the sky. The Main Ring is surrounded by a glow of \oiii\ emission and two low ionization halos. All three of these last features are visible as fossil radiation, material that was previously photoionized but is now shielded from the ionizing radiation of the central star. Hundreds of dark knots are best seen in extinction against the Main Ring's \oiii\ emission with the F502N filter, although many of them have low ionization arcs on the side of the knot that faces the central star. The dark knots generally point toward the central star, but do not have the radial symmetry of form seen in the Helix Nebula \citep{ode05}.

Since the polar axis of \ngc\ is pointed within about 10\arcdeg\ of the observer, the Main Ring is viewed nearly in the plane of the sky and tangential velocities can be determined. These tangential velocities can be used to understand the structure and distance of the nebula.  The Ring Nebula is also a prime target for addressing the important question of the relative motion of the nebular gas and the dark knots.

As part of HST program GO 12309 high signal-to-noise ratio images were made in the strongest emission lines of the entire nebula using the Wide Field Camera 3 (WFC3) that replaced the WFPC2 during the last servicing mission. These images were used to create the 3-D model discussed in O'Dell et al. (2013a). Images were also with smaller field-of-view diagnostic filters, and these were used to determine the physical conditions (ionization, temperatures, densities) within the nebula \citep{ode13b}.  The WFC3 pipeline-processed images sample better (40 mas per pixel) the point-spread-function of the HST and thus provide the highest resolution images of \ngc. As a new camera, no earlier observations exist of \ngc. However, one can accurately use the earlier WFPC2 images as first epoch images because the pipeline-processed images extracted from the HST archives have had an accurate distortion-correction applied, thus rendering the early and recent images geometrically ``flat''.  The geometric corrections step is called ``AstroDrizzle'' and is applied to both the old and new images. It is only necessary to scale the WFPC2 images to the same pixel size as the WFC3 images, align the central star images, and rotate them into alignment. One then has astrometrically useful paired images whose alignment accuracy should be about 0.2 WFC3 pixels. Not only are these recent images of higher spatial resolution, they offer the longer time base of 12.925 years when combined with the first images made in program GO 7632. 

In this paper we describe the method of aligning and measuring the image pairs (\S\ \ref{sec:combining}), present the results for measurements of gaseous features in \nii\ and the dark knots in F502N (\S\ \ref{sec:zsq} and \ref{sec:mag}), and discuss what this tells us about \ngc\ (\S\ \ref{sec:disc}).

\section{The Images}
\label{sec:images}

The images used in this study were obtained in the HST programs GO 7632 (WFPC2, 1998-10-16) and GO 12309 (WFC3, 2011-09-19).  Multiple exposures were made at each epoch, thus allowing cosmic ray event correction. The images in filter F502N isolate well the 500.7 nm line of \oiii, and the images in filter F658N isolate well the 658.3 nm line of  \nii. The \nii\ emission arises from a thin layer close to the main ionization front of the nebula \citep{ode13a}. In contrast, the \oiii\ emission arises from an extended region lying between the \nii\ emitting layer and the HeII core in the center of the nebula \citep{ode13a}. This means that the F658N images provide a means of tracing the motion of gas near the main ionization front, while the primary value of the F502N images lies in their ability to show the dark knots in detail.

\subsection{Combining the  WPC2 and WFC3 Images}
\label{sec:combining}

The first epoch GO 7632 AstroDrizzle images were downloaded from the HST data archive and put into the same pixel scale as the second epoch GO 12309 AstroDrizzled by using the IRAF \footnote{IRAF is distributed by the National Optical Astronomy Observatories, which is operated by the Association of Universities for Research in Astronomy, Inc.\ under cooperative agreement with the National Science foundation.} task ``magnify'' and the scaling factor of 2.5262758. The scaling factor could not be determined from the few stars in both Ring Nebula images was obtained by comparing rich star fields in the Orion Nebula Cluster imaged with the WFPC2 in program GO 5085 and with the WFC3 in program GO 12543. The two sets of images were aligned with IRAF task ``geomap/rotate'' using the four stars available. The alignment accuracy of the central star is about  3 mas,  and the determination accuracy  of the rotation is about 0.015\arcdeg.  

\begin{figure}
\epsscale{1.0}
\plotone{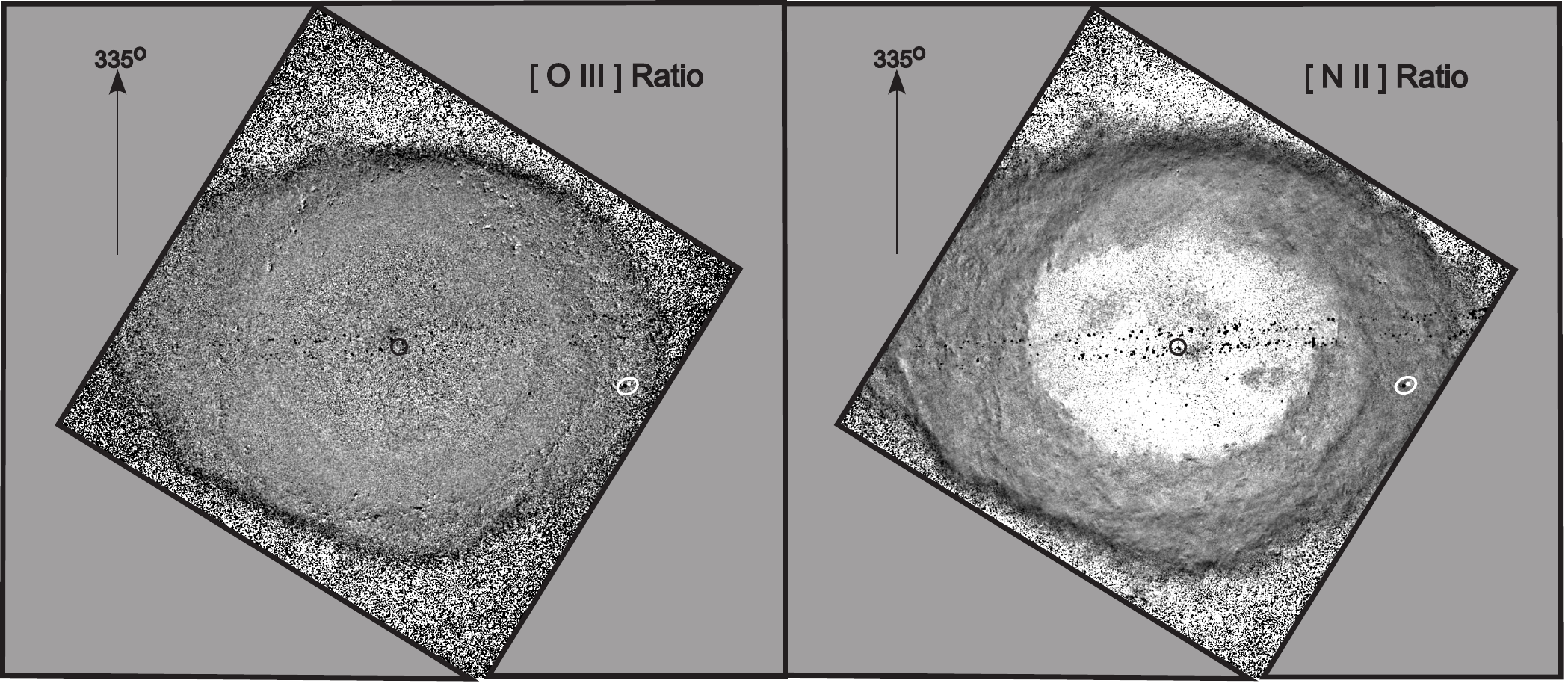}
\caption{These two images are both 120\arcsec\ by 104\arcsec\  and have been cropped to show only the region common to both the first epoch and second epoch images. They show the ratio of the first epoch divided by the second epoch images in the F502N and F658N filters.  The method of their derivation is described in \S\ \ref{sec:images}. For bright objects, the F658N ratio image shows motion of objects in the direction of the dark edge. For the dark knots, the ratio image  in F502N shows motion of objects in the direction of the bright edge. The nearly horizontal double band of small dark dots  across the middle of the image are caused by cosmic ray artifacts in the region of the WFC3 image for which adequate cosmic ray artifact correction could not be made. The location of the central star is indicated by a black circle. Note that the high proper motion star enclosed in the white oval was not used in aligning the two epoch images. 
\label{fig:ratios}}
\end{figure}

The aligned images were scaled to the same signal level, and the WFC3 image was blurred to the same stellar image point-spread-function as the WFPC2 image using the IRAF task ``gauss''. A ratio of the first and second epoch images was made, with the second epoch as the denominator. These ratio images revealed motions that had occurred during the 12.925 yr interval between the observations. For bright objects, the F658N ratio image shows motion of objects in the direction of the dark edge. For the dark knots, the ratio image  in F502N shows motion of objects in the direction of the bright edge. These images are shown in Figure \ref{fig:ratios}. 

\subsection{Measuring the Motion of Features with a Least-Squares Method}
\label{sec:zsq}

\begin{figure}
\epsscale{1.0}
\plotone{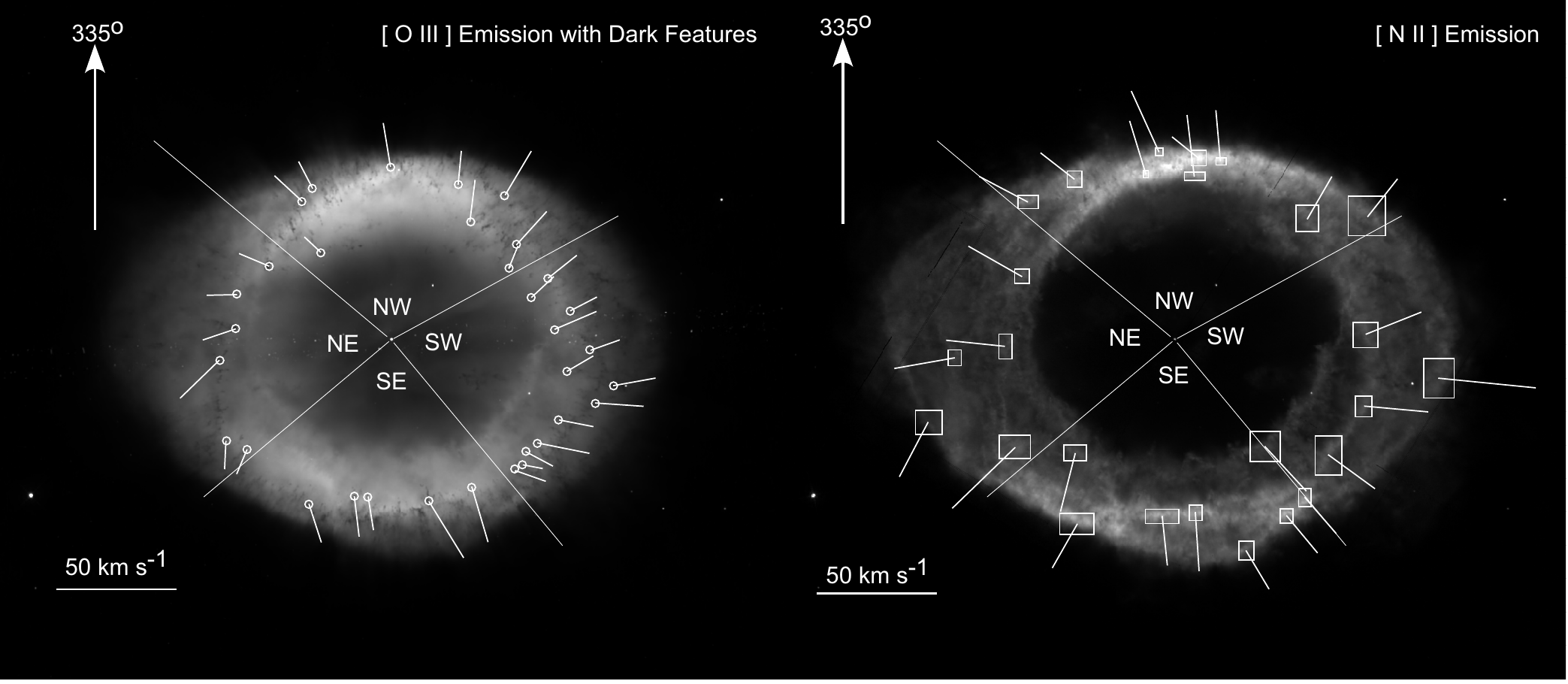}
\caption{The same fields of view as in Figure \ref{fig:ratios} are shown. The motions determined by the ZSQ method described in \S\ \ref{sec:images} are shown as vectors. At an assumed distance of 740 pc, a motion of one ACS pixel corresponds to a tangential velocity of 10.9 \kms. The velocity scale for the vectors is shown on both panels.  The small samples used for determination of the dark knots' motions are shown as open circles in the F502N image dominated by background \oiii\ emission and the larger samples used for determination of the bright \nii\ features' motions are shown as open boxes in the \nii\ image. 
\label{fig:motions}}
\end{figure}

The ratio images were used as a guide in determining samples to be measured using a least-squares code developed by Hartigan et al. (2001) from an approach originated by Currie et al. (1996) that compares incrementally shifted images. The motion is assumed to be correct when there is a minimum difference between the reference and shifted image.
In this paper we call this method the ZSQ method. The results for both small dark features seen in the F502N images and large bright features seen in the F658N images are shown in Figure~\ref{fig:motions}. Results for individual determinations of \mut\ are given in Table 1 and Table 2.  After recognizing patterns in the tangential velocities we grouped the \mut\ values into quadrants within the Main Ring as shown in Figure~\ref{fig:motions}. The averages within the quadrants are presented in Table 3.

It is possible to approximate the uncertainties of the measurements. The ZSQ gives much smaller uncertainties than indicated by a comparison of \mut\ for two nearby objects. Excluding the F502N dark features in the SW quadrant (c.f. \S \ref{sec:disc}), the average motion of all features is 2.12$\pm$0.33 ACS pixels. Over the time interval of the observations, this corresponds to \mut\ = 6.56 \masyr. If the alignment uncertainty is that of the central star image ( 3 mas or 0.08 pixels), that component of the uncertainty of the magnitude of \mut\ is only 0.2 \masyr\ and an uncertainty in the direction of motion of 2.2\arcdeg. The average distance of this set of samples is 27.5\arcsec\  (688 pixels). The uncertainty in the rotation between the two sets of images (0.015\arcdeg) would cause the tips of the features to be displaced by 0.18 pixels, which would produce an error in the direction of motion of 4.9\arcdeg. These two considerations mean that uncertainty in the alignment propagates into small uncertainties in the derived magnitude and direction of the motion of features in the nebula. Typically, the uncertainties in the ZSQ solutions are about 0.05 pixels, which corresponds to a 3\%\ uncertainty in the magnitude of \mut\ and an uncertainty of about 1.5\arcdeg.
The significant differences in the magnitude and direction of \mut\ for nearby objects in the Main Ring are likely to be real and can represent significant small-scale fluctuations in the gas flow of the Main Ring. An alternative explanation may be that two adjacent lines of sight are dominated by radiation from very different parts of the 3-D Main Ring.

\subsection{Measuring Motions with the Magnification Method}
\label{sec:mag}

\begin{figure}
\epsscale{1.0}
\plotone{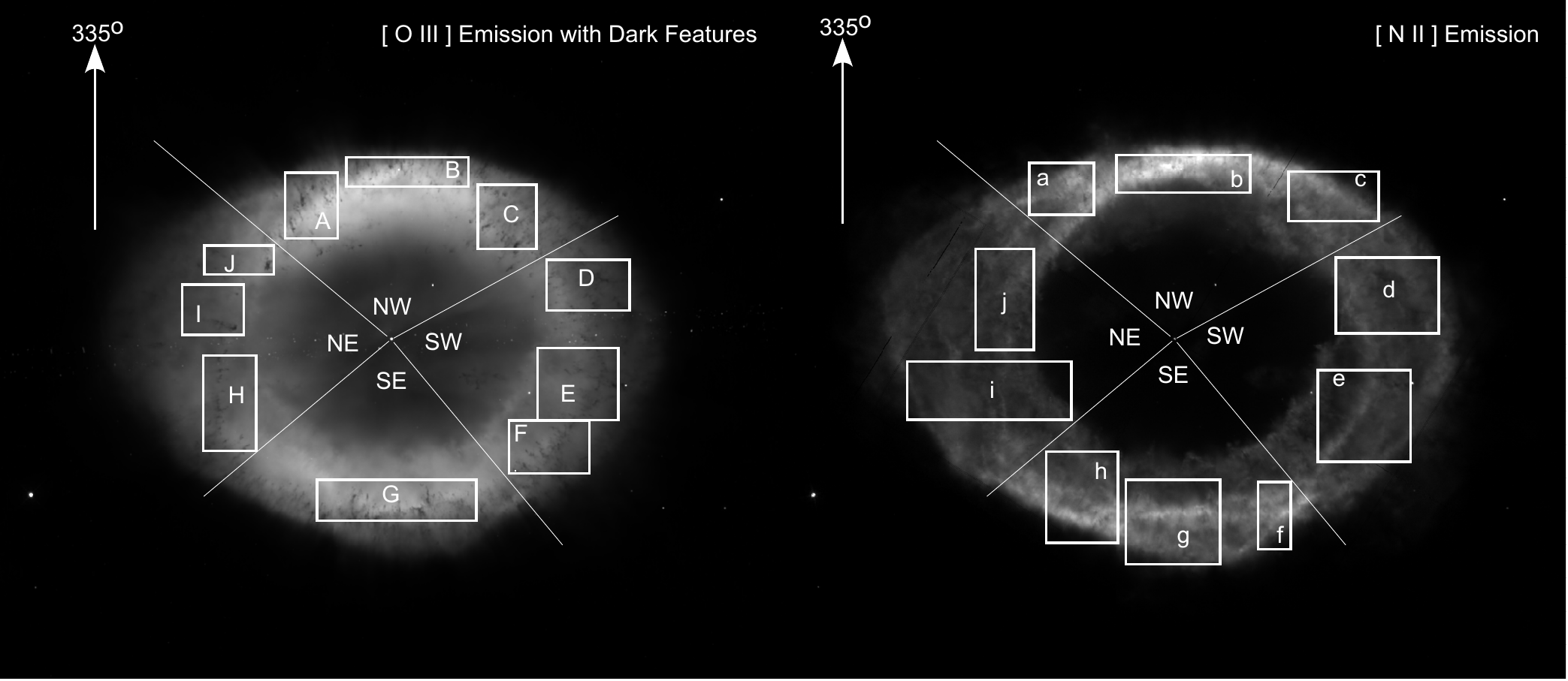}
\caption{The same fields of view as in Figure \ref{fig:ratios} are shown. The areas used in determining the optimum magnification are designated by heavy closed rectangles. 
\label{fig:mag}}
\end{figure}

We adopted a variation of the magnification method of determining motions. The underlying assumption of the method is that the tangential motions are homologous. 
When a correctly magnified first epoch image is aligned with the second epoch image (using the central star) and a ratio of the two is made, a flat image of constant value should result. 
However, if the expansion scale factor is not constant in all parts of the nebula, the employing a single optimum magnification (where the ratio of images becomes constant) will not represent fully what is happening. 
Recall from \S\ \ref{sec:6720history} that E and the magnification (M) are related by E=(M-1)$\times$10$^{3}$  when 
E is conveniently expressed in units of \Escale. 
We determined the optimum magnification and E for individual regions within the image of the Main Ring, as shown in Figure~\ref{fig:mag}. 
 The F502N samples were selected from where the nebular background varied little and contained numerous dark knots. The results should represent the motion of the dark knots. The F658N samples were selected from regions showing structure in the \nii\ image. The results should indicate motion of ionization fronts aligned with the observer's line-of-sight. 
 
Several steps were used to derive the optimum magnification. Initially, the first and second epoch images were scaled to the same signal level.  The first epoch images were magnified in steps of 0.00005. The central star was aligned on the resultant images, and a ratio of the two images was made for each magnification. Finally, the IRAF task ``imstat'' was applied to establish the standard deviation (STDDEV) of the signal within the sampled region. There was always a difference between the two images. The STDDEV was largest when the magnification was furthest from optimum, because the ratio-image was the result of comparing two distinct images, each with its own characteristics. However, as the magnification factor reached the optimum value, the STDDEV 
reached a minimum, then increased at values larger than the optimum magnification. The STDDEV was never zero because the signal-to-noise ratio of the first and second epoch images was not the same. Imperfect matching of the image resolution would also contribute to a non-zero STDDEV value. The images had been approximately matched to the same point-spread-function, but subtle differences in fine-structure would exist and contribute to a non-zero STDDEV even at the optimum magnification.The results of the optimum magnification and the expansion scale factor  determinations are given in Table 4 and Table 5. The group averages of the magnification results are given in Table 6.

\section{Discussion}
\label{sec:disc}

\subsection{Large Scale Properties of the Ring Nebula's Main Ring}
The results of the analysis of the motions was made in terms of average values of E. The quadrants we adopt are primarily meant to recognize regions of different motion, but also they agree with the natural divisions suggested by the most current 3-D model \citep{ode13a}.  A remarkable symmetry of features appears in opposite quadrants, which indicates similarity of properties. This justifies grouping the results into the quadrants along the major (NE+SW) and minor (NW+SE) axes. We adopt as the average of the ZSQ and magnification methods $<$E(\nii ,minor)$>$ = 0.22$\pm$0.04, $<$E(\nii ,major)$>$ = 0.27$\pm$0.05, $<$E(F502N,minor)$>$ = 0.20$\pm$0.03, and $<$E(F502N,major)$>$ = 0.18$\pm$0.02.

Our measurements of \mut\ are limited to the Main Ring of the nebula and include no measurements of inner halo or outer halo features.  O'Dell et al. (2007) first established and O'Dell et al. (2013a) confirmed the ellipticity of the Main Ring. The nebula's  equatorial concentration of material is elliptical in form, rather than the apparent ellipticity being the result of a circular ring seen at an angle. The small difference of the major and minor axes $<$E(\nii)$>$  values is consistent with the intrinsically elliptical equatorial concentration of the Main Ring. The minor axis material is expanding at a slower rate either because it was subject to less force during the phase when the Main Ring was being formed, or the over-burden of outer neutral material is greater there.

\subsection{Relative Velocities of the Dark Knots and the Nebular Gas}
Systematic differences of the averaged E values in \nii\ and F502N are seen to exist along both axes. This difference is greatest along the major axis, where $<$E(\nii)$>$ = 0.27$\pm$0.05 and $<$E(F502N)$>$ = 0.18$\pm$0.02. The difference along the minor axis is much more uncertain, with $<$E(\nii)$>$ = 0.22$\pm$0.04 and $<$E(F502N)$>$ = 0.20$\pm$0.03. A smaller value of E(F502N, dark knots) as compared with E(\nii, bright gas) indicates that the dark knots do not expand as fast as the ionized gas in which they exist. 

Two families of models are used to explain the origin of the dark knots found in PNe. The first \citep{dyson89} argues that the knots are condensations of dust and gas that originated in the extended atmosphere of the central star while it was still at the top of the Asymptotic Giant Branch. The similarity of the E values found today argue against this origin. It would be highly unlikely that knots sufficiently dense to survive the first phases of shell ejection would receive a  radial acceleration this similar to that of the low density gas. The second set of models \citep{cap73,vis83,gar06} invoke formation of the knots much later in the life of the nebula. Considerable observational evidence \citep{ode02, ode04} indicates that the knots arise near the main ionization front.  Our results show that the knots, once formed, appear to lag behind the motion of the surrounding gas, and this feature must be factored into theories of their formation and calculations of the subsequent properties. 

Previously  Meaburn et al. (1998) argued that the system of well-defined dark knots in the Helix Nebula are expanding at a rate of about one-half that of the nebular gas. That conclusion was drawn from study of the radial velocities of the small crescents of ionized gas formed on the side of the knots facing the central star. The primary concentration of gas in the Helix Nebula is in an equatorial ring lying almost in the plane of the sky \citep{mea98,ode04}, much like \ngc. Expansion motions are nearly in the plane of the sky, meaning that results determined from radial velocities are inherently less valuable. A study of the tangential velocities should be more accurate for addressing the question of the relative motion of the knots. However, it appears that the study of tangential velocities of \ngc\ and the radial velocities in the Helix Nebula come to the same conclusion: the knots are expanding more slowly and are decoupled from the gas motion.

\subsection{The Distance to \ngc}
\label{sec:distance}
\newcommand\Rfactor{\ensuremath{\mathcal{R}}}
A distance (D) for \ngc\ can be obtained from an accurate knowledge of its dynamical properties. For homologous expansion, the relation that applies is 
D(pc) = (k$\rm _{1} \times \Rfactor \times cos\theta$)/(4.74 E), where k$\rm _{1}$ is the scale factor relating the spatial expansion velocity (\kms) to the true (not projected) distance to the central star (expressed as \arcsec),  \Rfactor{} is the pattern velocity (the ratio of the ionization front velocity compared with the gas velocity), $\theta$ is the angle of the equatorial plane of the nebula to the sky (\arcdeg), and E is the expansion scale ratio. Adopting the values in O'Dell et al. (2007) of k$\rm _{1}$ = 0.65 (\kms / ") and \(\Rfactor = 1.34\), the relation becomes D(pc) = 179$\times$E$^{-1}$, where E is in \Escale. From this work we adopt the average of the major and minor axes for \nii\ (E=0.25 \Escale) and find D = 716 pc. It is difficult to identify an uncertainty in this distance because it is dependent upon the validity of the dynamical model used. The key uncertainties are in the determination of  k$\rm _{1}$ \citep{OSH} and \Rfactor{} \citep{OHS}. The resulting uncertainty in the distance is probably about 30\%. The derived distance should  be compared with the astrometric distance of 700$\pm ^{450} _{200}$ pc \citep{har07}.  Our rounded value of 720 pc is the best distance to be adopted. This is only a small change from the previously recommended value of 740 pc \citep{OHS}. 

\subsection{Timescales}

The reciprocal of the expansion ratio has the dimension of years and represents the time since the onset of the assumed homologous expansion. Adopting the average of E = 0.25 (\S \ref{sec:distance}) gives 4000 yrs since onset of the expansion. This method of calculation produces an overly simplified dynamical age and only has the meaning as a measure of when the basic dynamics properties of the expanding nebular shell were established under the assumption that the size at that point was small.

The current characteristics of the central star are (log~L/\lsol) = 2.3$\pm$0.2 and log~T $\approx$ 5.1$\pm$0.02) \citep{OSH} and comparison with the theoretical path for a collapsing star of about M = 0.62 \msol \citep{blo95} indicates that the time since onset of stellar collapse is about 5000 yrs, i.e. similar to the dynamic age. 

\section{Summary and Conclusions}

We have been able to utilize early HST WFPC2 images of \ngc\ in combination with recent WFC3 images to determine tangential motions of \nii\ emission features near the main ionization front of this ionization-bounded nebula. We could also determine the tangential velocities of the dark knots seen as extinction of the diffuse nebular \oiii\ emission.

Well defined radial expansion patterns were found in both the \nii\ emission-line features and the dark knots. The results from a second method of determination of the motions under the assumption of homologous expansion are in excellent agreement with the results from measurement of individual features. 

The expansion scale ratio E that characterizes the scale of motions within the nebula is not the same on both axes of the Main Ring.  It is larger for the major axis, a result consistent with the inherent ellipticity of the Main Ring. The apparent ellipticity of the Main Ring is primarily due to its inherent shape, rather than being the result of a circular object being observed at an angle to the plane of the sky. The suggestions of spiral structure are the results of the asymmetries of the boundaries of the Main Ring and the lobes \citep{ode13a}.

The reciprocal of the expansion scale factor indicates that the dynamic age of \ngc\ is about 4000 yrs. Combining the new results for the expansion factor and the previously established spatial expansion velocity scale factor determined from radial velocities has allowed the calculation of a distance to \ngc\ of 720 pc, with a likely uncertainty of 30 \% .

 \acknowledgments
We are grateful to David Thompson of the Large Binocular Telescope Observatory for providing copies of his unpublished LBTO \htwo\ data taken with the LUCI1 instrument and to Zoltan Levay of the Space Telescope Science Institute  for a fruitful discussion on this study.

 GJF acknowledges support by NSF (0908877; 1108928; and 1109061), NASA (10-ATP10-0053, 10-ADAP10-0073, and NNX12AH73G), JPL (RSA No 1430426), and STScI (HST-AR-12125.01, GO-12560, and HST-GO-12309).  MP received partial support from CONACyT grant 129553. WJH acknowledges financial support from DGAPA--UNAM through project PAPIIT IN102012. CRO's participation was supported in part by HST programs GO 12309 and GO 12543 . 

{\it Facilities:} \facility{HST {(WFC3)}}


\begin{deluxetable}{lccclc}
\tabletypesize{\scriptsize}
\tablecaption{ZSQ Method Motions for \nii\ Bright Features}
\tablewidth{0pt}
\tablehead{
\colhead{Sample--PA* (\arcdeg)} &
\colhead{Sample--Distance (\arcsec)} &
\colhead{\mut\ (\masyr)} &
\colhead{Vector--PA* (\arcdeg)} &
\colhead{E~(\Escale)} &
\colhead{Quadrant}}
\startdata
021.9     &          30.8     & ~6.51  &     037     &    0.211  & NW \\
007.1     &           28.9    &  ~5.11  &     027&         0.177&--\\
344.8     &         25.7   &    ~6.66    &   352     &    0.259 &--\\
339.7     &         28.8  &    ~7.97  &     359   &      0.277&--\\
328.1    &          24.8  &    ~7.36   &    342    &     0.297&--\\
327.1     &         27.8  &    ~4.14   &    026    &     0.167&--\\
320.4     &         28.2   &   ~6.08    &   340   &      0.216&--\\
287.2     &         27.4   &   ~5.85    &   305   &      0.213&--\\
277.4     &         35.1  &   ~ 5.75    &   297   &      0.164&--\\

246.5   &           29.3   &   ~7.12    &   267    &     0.243  & SW \\
236.5    &          40.7 &    11.63  &     239   &      0.286&--\\
225.5    &          30.9   &   ~7.63   &    240   &      0.247&--\\
208.0   &           29.4   &   ~6.88   &    209   &      0.234&--\\
195.1    &          21.5  &    ~7.25  &     198     &    0.338&--\\

194.3    &          31.5   &  ~5.71   &    196    &     0.181   & SE \\
187.3     &         31.8  &   ~ 5.79   &    195    &     0.182&--\\
173.5     &         34.1   &   ~5.35    &   186   &      0.157&--\\
161.8     &         26.7   &   ~7.01   &    159   &      0.263&--\\
150.9     &         27.2  &    ~5.91    &   161   &      0.217&--\\
127.3    &          32.0   &  ~ 6.03    &   126   &      0.189&--\\
114.0    &          23.2   &   ~7.17   &    141    &     0.309&--\\

099.0    &           29.5  &   10.41    &   109   &      0.353   & NE \\
083.5    &           39.9   &   ~7.34   &    127    &     0.184   &--\\
069.9    &           33.9  &    ~7.29   &    075    &     0.215&--\\
067.3   &            26.1   &   ~7.00    &   059   &      0.268&--\\
042.5     &          25.3   &   ~7.33    &   035   &      0.290&--\\
\enddata
\tablecomments{~*PA means the Position Angle, measured counter-clockwise from north.}
\end{deluxetable}

\begin{deluxetable}{lccclc}
\tabletypesize{\scriptsize}
\tablecaption{ZSQ Method Motions for F502N Dark Features}
\tablewidth{0pt}
\tablehead{
\colhead{Sample--PA* (\arcdeg)} &
\colhead{Sample--Distance (\arcsec)} &
\colhead{\mut\ (\masyr)} &
\colhead{Vector--PA* (\arcdeg)} &
\colhead{E~(\Escale)} &
\colhead{Quadrant}}
\startdata
014.4      &         16.3   &   5.45 &     045     &       0.334    & NW  \\
007.8           &     24.2   &   7.16    &  040       &     0.295&--\\
002.4           &     25.5    &  5.86    &  032        &    0.230&--\\
334.3      &        25.8  &    6.63   &   013      &      0.257&--\\
310.0       &       25.4  &    4.92   &   010     &       0.194&--\\
300.9        &      21.8  &    5.63  &    002      &      0.258&--\\
296.7       &       28.0  &    5.23   &   336      &      0.188&--\\
280.1       &       24.1  &    4.01   &   329      &      0.166&--\\
274.1      &        21.1  &    3.40    &  013        &    0.161&--\\
266.3       &       25.7  &    2.73    &  331       &     0.089   & SW \\
261.6      &        22.4  &    2.54  &    347      &      0.113&--\\
252.6      &        28.1  &    1.64   &   247      &      0.058&--\\
248.4       &       25.1  &    2.81   &   296      &      0.112&--\\
240.0       &       30.6  &    1.26   &   323      &     0.041&--\\
233.6       &       28.0  &    1.80   &   336        &    0.064&--\\
233.1        &      34.8  &    2.05    &  271       &     0.059&--\\
227.6     &         32.8  &    2.61   &   237      &      0.080&--\\
219.2      &        28.6   &   1.18   &   202      &      0.041&--\\
209.6      &        27.5   &    3.28    &  222        &    0.119&--\\
205.5      &        26.7 &    1.76   &   169      &      0.066&--\\
201.3     &        27.9  &    0.93   &   093       &     0.033&--\\
198.7      &        27.5  &    1.34  &    173       &     0.049&--\\
183.2    &         25.7  &    6.79    &  145       &     0.264   &SE \\
168.7     &         26.1  &    6.77   &   163      &      0.253&--\\
148.0     &         25.0  &    4.75   &   123        &    0.190&--\\
153.1     &         25.3   &   5.48    &  126     &       0.217&--\\
129.8      &        28.6  &    4.91   &   135     &       0.172&--\\
104.0       &       27.8   &   5.30    &  098       &     0.191   &NE  \\
098.0       &        29.8   &   4.77    &  110      &      0.160&--\\
073.5       &        26.1   &   9.09   &   095      &      0.348&--\\
062.5       &        23.3   &   7.26   &   076      &      0.312&--\\
048.6        &       24.8   &   6.81   &   066      &      0.274&--\\
034.9        &       21.2   &   6.96    &  053      &      0.329&--\\
\enddata
\tablecomments{~*PA means the Position Angle, measured counter-clockwise from north.}
\end{deluxetable}

\begin{deluxetable}{lcc}
\tabletypesize{\scriptsize}
\tablecaption{Averaged  ZSQ Motions Grouped by Quadrants}
\tablewidth{0pt}
\tablehead{
\colhead{Region} &
\colhead{Average \mut\ (\masyr)} &
\colhead{Average Expansion Scale Factor (\Escale)} }
\startdata
\nii\ Bright~ NW & 6.2$\pm$1.2 & 0.22$\pm$0.05\\
\nii\ Bright~SW & 8.0$\pm$2.0  & 0.27$\pm$0.04\\
\nii\ Bright~SE &  6.1$\pm$0.7 & 0.21$\pm$0.05\\
\nii\ Bright~NE  & 7.9$\pm$1.4 & 0.26$\pm$0.07\\
\nii\ Bright~NW+SE &  6.2$\pm$0.9 & 0.22$\pm$0.05\\
\nii\ Bright~NE+SW & 8.0$\pm$1.6 & 0.27$\pm$0.05\\
F502N~Dark~NW & 4.4$\pm$1.2 & 0.19$\pm$0.04\\
F502N~Dark~SW & 4.3$\pm$1.1 & 0.15$\pm$0.04\\
F502N~Dark~SE & 5.7$\pm$1.6 & 0.22$\pm$0.06\\
F502N~Dark~NE & 4.1$\pm$1.1 & 0.17$\pm$0.05\\ 
F502N~Dark~NW+SE & 4.9$\pm$1.4 & 0.20$\pm$0.05\\
F502N~Dark~NE+SW & 4.2$\pm$1.1 & 0.16$\pm$0.04\\
\enddata
\end{deluxetable}

\begin{deluxetable}{cllc}
\tabletypesize{\scriptsize}
\tablecaption{Magnification Method Results for \nii\ Bright Samples}
\tablewidth{0pt}
\tablehead{
\colhead{Sample} &
\colhead{Optimum Magnifier} &
\colhead{E~(\Escale)} &
\colhead{Quadrant}}
\startdata
a      &      1.00275$\pm$0.00025 &  0.21$\pm$0.02 & NW \\
b      &      1.0035~$\pm$0.00025 &  0.27$\pm$0.02 & NW \\
c      &      1.00288$\pm$0.00025 &  0.22$\pm$0.02 & NW \\
d      &      1.00275$\pm$0.00025 &  0.21$\pm$0.02 & SW \\
e      &      1.00425$\pm$0.00025 &  0.33$\pm$0.02 & SW \\
f       &      1.0025~$\pm$0.00025 &  0.19$\pm$0.02 & SE \\
g      &      1.00325$\pm$0.00025 &  0.25$\pm$0.02 & SE \\
h      &      1.0025~$\pm$0.00025 &  0.19$\pm$0.02 & SE \\
i       &      1.00425$\pm$0.00025 &  0.33$\pm$0.02 & NE \\
j       &      1.00325$\pm$0.00025 &  0.25$\pm$0.02 & NE \\
\enddata
\tablecomments{~*Sample regions are shown in Figure~\ref{fig:mag}.}
\end{deluxetable}

\begin{deluxetable}{cllc}
\tabletypesize{\scriptsize}
\tablecaption{Magnification Method Results for F502N Samples}
\tablewidth{0pt}
\tablehead{
\colhead{Sample} &
\colhead{Optimum Magnifier} &
\colhead{E~(\Escale)} &
\colhead{Quadrant}}
\startdata
A      &      1.0025~$\pm$0.00025 &  0.19$\pm$0.02 & NW \\
B     &      1.00275~$\pm$0.00025 &  0.21$\pm$0.02 & NW \\
C     &      1.00225$\pm$0.00025 &  0.17$\pm$0.02 & NW \\
D      &      1.00225$\pm$0.00025 &  0.17$\pm$0.02 & SW \\
E      &      1.00225$\pm$0.00025 &  0.17$\pm$0.02 & SW \\
F       &      1.0025~$\pm$0.00025 &  0.19$\pm$0.02 & SW \\
G     &      1.00275$\pm$0.00025 &  0.21$\pm$0.02 & SE \\
H      &      1.00275~$\pm$0.00025 &  0.21$\pm$0.02 & NE \\
I      &      1.0025~$\pm$0.00025 &  0.19$\pm$0.02 & NE \\
J       &      1.00225$\pm$0.00025 &  0.17$\pm$0.02 & NE \\
\enddata
\tablecomments{~*Sample regions are shown in Figure~\ref{fig:mag}.}
\end{deluxetable}

\begin{deluxetable}{lcc}
\tabletypesize{\scriptsize}
\tablecaption{Averaged  Magnification Method  Results  Grouped by Quadrants}
\tablewidth{0pt}
\tablehead{
\colhead{Region} &
\colhead{1-Average Optimum Magnifier} &
\colhead{Average Expansion Scale Factor (\Escale)} }
\startdata
\nii\ Bright~ NW & 3.04$\pm$0.40 & 0.23$\pm$0.03\\
\nii\ Bright~SW & 3.50$\pm$0.70  & 0.27$\pm$0.06\\
\nii\ Bright~SE &  2.80$\pm$0.35 & 0.21$\pm$0.04\\
\nii\ Bright~NE  & 3.75$\pm$0.5 & 0.29$\pm$0.04\\
\nii\ Bright~NW+SE &  2.73$\pm$0.30 & 0.22$\pm$0.03\\
\nii\ Bright~NE+SW & 3.62$\pm$0.75 & 0.28$\pm$0.06\\
F502N~Dark~NW & 2.50$\pm$0.25 & 0.19$\pm$0.02\\
F502N~Dark~SW & 2.33$\pm$0.14 & 0.18$\pm$0.01\\
F502N~Dark~SE & 2.75$\pm$0.25 & 0.21$\pm$0.02\\
F502N~Dark~NE & 2.5$\pm$0.25 & 0.19$\pm$0.02\\ 
F502N~Dark~NW+SE & 2.60$\pm$0.22 & 0.20$\pm$0.02\\
F502N~Dark~NE+SW & 2.42$\pm$0.20 & 0.18$\pm$0.02\\
\enddata
\end{deluxetable}

\end{document}